\documentclass[twocolumn]{revtex4}
\usepackage{epsfig}
\usepackage{amssymb}
\usepackage{amsmath}
\usepackage{amsfonts}
\usepackage{graphicx}
\usepackage{mathrsfs}
\usepackage[dvips]{color}
\usepackage{multirow}

\newcommand{\R}{\mathbb{R}}

\newcommand{\fz}{\mathfrak{z}}

\newcommand{\cH}{\mathcal{H}}

\newcommand{\cO}{\mathcal{O}}
\newcommand{\cP}{\mathcal{P}}

\newcommand{\be}{\begin{equation}}
\newcommand{\ee}{\end{equation}}
\newcommand{\bea}{\begin{eqnarray}}
\newcommand{\eea}{\end{eqnarray}}

\newcommand{\ed}{\end{document}}

\newcommand{\bi}{\begin{itemize}}
\newcommand{\ei}{\end{itemize}}

\newcommand{\bce}{\begin{center}}
\newcommand{\ece}{\end{center}}

\newcommand{\sG}{\mathscr{G}}

\begin{document}
\title{Nonlinear Spectral Singularities for Confined Nonlinearities}
\author{Ali~Mostafazadeh}
\affiliation{Department of Mathematics, Ko\c{c}
University, Sar{\i}yer 34450, Istanbul, Turkey}

\begin{abstract}
We introduce a notion of spectral singularity that applies for a general class of nonlinear Schr\"odinger operators involving a confined nonlinearity. The presence of the nonlinearity does not break the parity-reflection symmetry of spectral singularities but makes them amplitude-dependent. Nonlinear spectral singularities are, therefore, associated with a resonance effect that produces amplified waves with a specific amplitude-wavelength profile. We explore the consequences of this phenomenon for a complex $\delta$-function potential that is subject to a general confined nonlinearity.\\[3pt]
\hspace{6.2cm}{Pacs numbers: 03.65.Nk, 42.25.Bs, 42.65.-k,
24.30.Gd}
\end{abstract}

\maketitle

\noindent\emph{Introduction}.---  A spectral singularity of a complex scattering potential is a mathematical concept introduced and studied by mathematicians for more than half a century \cite{naimark,guseinov-2009}. This concept that applies only for non-real scattering potentials, was recently shown to have an intriguing physical interpretation \cite{prl-2009}: It corresponds to the energy of a scattering state whose reflection and transmission coefficients diverge. This observation has motivated identifying spectral singularities with certain zero-width resonances and led to the study of their physical implications \cite{prl-2009,oss,ss-phys}. In optics, a spectral singularity gives rise to lasing at the threshold gain \cite{pra-2011a}. Its time-reversal corresponds to a coherent perfect absorption (CPA) of light \cite{longhi-phys-10} that is also called anti-lasing. The observation of the latter reported in \cite{antilasing} may be considered as an experimental evidence for the physical relevance of spectral singularities.

Once one creates a spectral singularity in an optically active material \cite{prl-2009,oss}, it begins amplifying the background noise and functions as a laser. One may argue that a realistic treatment of this phenomenon should also take into account the nonlinearities associated with the production of high-intensity radiation in the gain region. This provides our main physical motivation for the study of the meaning and behavior of spectral singularities for nonlinear operators, a task that to the best of our knowledge has not been considered previously \cite{guseiynov}. The search for an appropriate nonlinear extension of the notion of a spectral singularity is particularly important not only because it has been an open mathematical problem for several decades, but also because it can pave the way for the discovery of the nonlinear analogs of the interesting physical phenomena such as threshold lasing \cite{pra-2011a}, antilasing \cite{antilasing}, and CPA-lasers \cite{longhi-phys-10} in areas other than optics.

Motivated by the physical meaning of a spectral singularity of a linear Schr\"odinger operator  \cite{prl-2009}, in this letter we propose a definition for a spectral singularity of the nonlinear Schr\"odinger operators $\cH_\gamma$ of the form
    \be
    H_\gamma\psi(x):=-\psi''(x)+v(x)\psi(x)+\gamma\,\chi(x)f(|\psi(x)|,x)\psi(x),
    \label{nonlinear}
    \ee
where $v$ is a rapidly decaying complex scattering potential, $\gamma$ is a nonzero real coupling constant, $\chi(x):=1$ for $x\in[0,1]$ and $\chi(x):=0$ for $x\notin[0,1]$, and $f$ is a real-valued function. The presence of $\chi$ in (\ref{nonlinear}) shows that $H_\gamma$ involves a confined nonlinearity. A concrete example is the confined Kerr nonlinearity, with $f(|\psi(x)|,x):=|\psi(x)|^2$, that appears in the study of Bose-Einstein condensates \cite{rwk} and has well-known applications in optics.

The problem of introducing spectral singularities for nonlinear operators is plagued with severe mathematical difficulties associated with the proposing an appropriate definition for the spectrum and a suitable scattering theory for these operators. The simple idea of considering confined nonlinearities, that is mainly motivated by physical considerations \cite{lepri}, allows for circumventing these difficulties. As we show below this idea plays a central role in our ability for defining a useful notion of a nonlinear spectral singularity (NSS).

The time-independent nonlinear Schr\"odinger equation corresponding to (\ref{nonlinear}) is given by
    \be
    H_\gamma\psi(x)=k^2\psi(x),
    \label{NL-sch-eq}
    \ee
where $k$ is a complex number. It is easy to see that outside the interval $[0,1]$, (\ref{NL-sch-eq}) coincides with the linear time-independent Schr\"odinger equation;
    \be
     -\psi''(x)+v(x)\psi(x)=k^2\psi(x).
    \label{eg-va-eq}
    \ee

\noindent\emph{Jost Solutions and Nonlinear Spectral Singularities}.---
Consider the linear operator $H:=-\partial^2_x+v(x)$, whose continuous spectrum is $[0,\infty)$. The eigenvalue equation for $H$, i.e., (\ref{eg-va-eq}), admits the so-called Jost solutions $\psi_{k\pm}$ that fulfil the asymptotic boundary conditions: $\psi_{k\pm}(x)\to N_\pm e^{\pm ikx}$ as $x\to\pm\infty$, for some nonzero complex numbers $N_\pm$. In particular,
    \be
    \lim_{x\to\pm\infty}\left[\psi'_{k\pm}(x)\mp ik\psi_{k\pm}(x)\right]=0.
    \label{jost-a}
    \ee
The Jost solutions, $\psi_{k+}$ and $\psi_{k-}$, are the scattering solutions of  (\ref{eg-va-eq}) corresponding to incident waves from the left and right, respectively.\\[6pt]
\noindent {\emph{Definition}~1}: A positive real number $k^2$ is called a spectral singularity of $H$ if $\psi_{k\pm}$ are linearly-dependent \cite{guseinov-2009}, i.e., $\psi_{k-}\propto\psi_{k+}$.\\[6pt]
The following is a simple consequence of this definition.\\[6pt]
\noindent {\emph{Theorem}~1}: A positive real number $k^2$ is a spectral singularity of $H$ if and only if there is a solution $\psi_{k}$ of (\ref{eg-va-eq}) such that
$\lim_{x\to\pm\infty}e^{\mp ikx}\psi_{k}(x)$ exist as nonzero complex numbers, and $\psi_k$ satisfies
    \be
    \lim_{x\to\pm\infty}\Big[\psi'_{k}(x)\mp ik \psi_{k}(x)\Big]=0.
	\label{condi-nonlinear}
    \ee
We introduce a notion of spectral singularity for the nonlinear operator (\ref{nonlinear}) by promoting Theorem~1 to a definition.\\[6pt]
{\emph{Definition}~2}: A positive real number $k^2$ is said to be a spectral singularity of $H_\gamma$, if there is a solution $\psi_{k}$ of (\ref{NL-sch-eq}) such that
$\lim_{x\to\pm\infty}e^{\mp ikx}\psi_{k}(x)$ exist as nonzero complex numbers, and $\psi_{k}$ satisfies (\ref{condi-nonlinear}).\\[6pt]
In what follows we use the term NSS for a spectral singularity of the nonlinear operator (\ref{nonlinear}).

In order to ensure that the physical interpretation of spectral singularities is left intact, we demand the existence of the Jost solutions of the nonlinear equation~(\ref{NL-sch-eq}). We identify  them with the solutions that satisfy the asymptotic boundary conditions (\ref{jost-a}), denote them by $\psi^{(\gamma)}_{k\pm}$, and keep using $\psi_{k\pm}$ for the Jost solutions of the linear equation (\ref{eg-va-eq}). Moreover, because the nonlinearity is confined to $[0,1]$ and $\psi^{(\gamma)}_{k\pm}$ are continuously differentiable,
    \be
    \begin{array}{c}
    \psi^{(\gamma)}_{k-}(x)=\psi_{k-}(x)~{\rm for}~x\leq 0,\\
    \psi^{(\gamma)}_{k+}(x)=\psi_{k+}(x)~~{\rm for}~~x\geq 1.
    \end{array}
    \label{e0}
    \ee
In particular,
    \bea
    \psi^{(\gamma)}_{k-}(0)=\psi_{k-}(0),~~~~
    \psi^{(\gamma)\prime}_{k-}(0)=\psi'_{k-}(0),
    \label{e1}\\
    \psi^{(\gamma)}_{k+}(1)=\psi_{k+}(1),~~~~
    \psi^{(\gamma)\prime}_{k+}(1)=\psi'_{k+}(1).
    \label{e2}
    \eea

We can view Eqs.~(\ref{e1}) and (\ref{e2}) as initial conditions for the differential equation (\ref{NL-sch-eq}). Solving the initial-value problem defined by (\ref{NL-sch-eq}) and (\ref{e1}) for $x>0$ gives  $\psi^{(\gamma)}_{k-}$ on $[0,\infty)$. Similarly, solving the initial-value problem defined by (\ref{NL-sch-eq}) and (\ref{e2}) for $x<0$, we find $\psi^{(\gamma)}_{k+}$ on $(-\infty,0]$. These together with (\ref{e0}) specify $\psi^{(\gamma)}_{k\pm}$ throughout $\R$. Note, however, that this procedure works provided that the above initial-value problems have global solutions. Indeed, because the nonlinearity is confined to $[0,1]$, it suffices to make sure that they have solutions on $[0,1]$. The spectral singularities of $H_\gamma$ are given by the values of $k^2$ for which at least one of the Jost solutions $\psi^{(\gamma)}_{k\pm}$ satisfy the conditions listed in Definition~2.\vspace{6pt}

\noindent\emph{{Potentials Vanishing Outside $[0,1]$.}}--- In Refs.~\cite{pra-2011b,pra-2012}, we consider linear spectral singularities (LSSs) of potentials that vanish outside $[0,1]$. This allows for making more definitive statements about the behavior of these spectral singularities. The same holds for the NSSs.

Suppose that $v(x)=0$ for $x\notin[0,1]$. Then, Eqs.~(\ref{e0}) and (\ref{e1})--(\ref{e2}) respectively take the form $\psi^{(\gamma)}_{k-}(x)=N_-e^{-i kx}$ for $x\leq 0$, $\psi^{(\gamma)}_{k+}(x)=N_+e^{i kx}$ for $x\geq 1$, and
    \bea
    \psi^{(\gamma)}_{k-}(0)&=&N_-,~~~~~~
    \psi^{(\gamma)\prime}_{k-}(0)=-i k N_-,
    \label{e11}\\
    \psi^{(\gamma)}_{k+}(1)&=&\tilde N_+,~~~~~~
    \psi^{(\gamma)\prime}_{k+}(1)=ik \tilde N_+,
    \label{e12}
    \eea
where $\tilde N_+:=N_+e^{i k}$. In order to determine $\psi^{(\gamma)}_{k-}(x)$ for $x>0$, we solve the initial-value problem defined by  (\ref{NL-sch-eq}) and (\ref{e11}) on $[0,1]$. The solution, that we denote by $\xi_{k}$, gives the value of $\psi^{(\gamma)}_{k-}(x)$ for $x\in[0,1]$, provided that it exists. For $x>1$, the general solution of (\ref{NL-sch-eq}) is a linear combination of plane waves. This observation together with the requirement that $\psi^{(\gamma)}_{k-}$ is continuously differentiable at $x=1$ give
    \be
    \psi^{(\gamma)}_{k-}(x)\!=\!
    \left\{\begin{array}{ccc}
    N_-e^{-ikx}&{\rm for}& x<0,\\[3pt]
    \xi_{k}(x)&{\rm for}& 0\leq x\leq 1,\\[3pt]
    \frac{e^{ik(x-1)}F_+(k)-
    e^{-ik(x-1)}F_-(k)}{2ik}
    &{\rm for}& x> 1.
    \end{array}\right.
    \label{jm}
    \ee
Here $\xi_k$ is the solution of  (\ref{NL-sch-eq}) on $[0,1]$ that satisfies
    \be
    \xi_k(0)=N_-,~~~~~~\xi'_k(0)=-ikN_-,
    \label{ini-1}
    \ee
and $F_{\pm}(k):=\xi'_k(1)\pm i k\xi_k(1)$. Similarly, we obtain
    \be
    \psi^{(\gamma)}_{k+}(x)=
    \left\{\begin{array}{ccc}
     \frac{e^{ikx}G_+(k)-e^{-ikx}G_-(k)}{2ik}
    &{\rm for}& x<0,\\[3pt]
    \zeta_{k}(x)&{\rm for}& x\in[0,1],\\[3pt]
    \tilde N_+ e^{ik(x-1)} &{\rm for}& x> 1,
    \end{array}\right.
    \label{jp}
    \ee
where $G_{\pm}(k):=\zeta'_k(0)\pm i k\zeta_k(0)$ and $\zeta_k$ is the solution of (\ref{NL-sch-eq}) on $[0,1]$ that fulfils the initial conditions: $\zeta_k(1)=\tilde N_+$ and $\zeta'_k(1)=ik \tilde N_+$.

Because $\psi^{(\gamma)}_{k+}$ and $\psi^{(\gamma)}_{k-}$ respectively correspond to the scattering states with incident wave from the left and right, we can use (\ref{jm}) and (\ref{jp}) to determine the left and right reflection and transmission amplitudes, $R$ and $T$. This gives
    \be
    \begin{aligned}
    &R^l=-\frac{G_-(k)}{G_+(k)}, &&T^l=\frac{2ik e^{-ik}\tilde N_+}{G_+(k)},\\
    &R^r=-\frac{e^{-2ik}F_+(k)}{F_-(k)}, &&T^r=-\frac{2ik e^{-ik}N_-}{F_-(k)},
    \end{aligned}
    \label{R-T}
    \ee
where the superscripts $l$ and $r$ stand for left and right, respectively. For $\gamma=0$, $G_\pm$ (resp.\ $F_\pm$) are proportional to $\tilde N_+$ (resp.\ $N_-$), and the latter drops out of (\ref{R-T}).

Having obtained the explicit form of $\psi^{(\gamma)}_{k\pm}$, we can impose the condition that they yield a NSS. Demanding that $\psi^{(\gamma)}_{k+}$ satisfies (\ref{condi-nonlinear}), we find $G_+(k)=0$. Similarly, imposing (\ref{condi-nonlinear}) on $\psi^{(\gamma)}_{k-}$ gives $F_-(k)=0$. Therefore, in view of (\ref{R-T}),  NSSs can again be interpreted as the energies of certain zero-width resonances \cite{prl-2009}.

The condition that either $\psi^{(\gamma)}_{k+}$ or $\psi^{(\gamma)}_{k-}$ gives rise to a NSS is equivalent to demanding that $\xi$ and $\zeta$ are solutions of the boundary-value problem defined on $[0,1]$ by (\ref{NL-sch-eq}) and the outgoing boundary conditions:
	\be
	\psi'_k(0)-ik\psi_k(0)=0,~~~~
	\psi'_k(1)+ik\psi_k(1)=0.
	\label{condi-nonlinear-2}
	\ee
It is easy to see that for potentials vanishing outside $[0,1]$ Eqs.~(\ref{condi-nonlinear-2}) are equivalent to (\ref{condi-nonlinear}). In particular, they are invariant under the parity ($\cP$) transformation: $x\to 1-x$. This is a manifestation of the fact that, similarly to the LSSs \cite{jpa-2012}, the $\cP$-symmetry of the boundary conditions (\ref{condi-nonlinear}) or (\ref{condi-nonlinear-2}) leads to an intrinsic $\cP$-symmetry of NSSs. This means that once the parameters of the system are tuned to realize a spectral singularity, it will amplify the background noise and begin emitting radiation of the same wavelength from both ends.

The main difference between LSSs and NSSs is that the resonance effect corresponding to the latter is intensity-dependent. The system amplifies an incident plane wave of negligible amplitude and produces outgoing waves of the same wavenumber $k$ and a particular $k$-dependent sizable amplitude. This intensity-dependent resonance effect may, for example, be used to devise a measurement scheme that determines the wavelength of an incident wave using the information about the intensity of the transmitted wave.\vspace{6pt}

\noindent\emph{{Complex $\delta$-function potential.}}--- Consider the potential
	\be
	v(x)=\fz\,\delta(x-a),
	\label{v=d}
	\ee
where $\fz$ is a complex coupling constant and $a\in(0,1)$. This potential supports a single LSS provided that $\fz$ is purely imaginary \cite{jpa-2006}. For this choice of $v$ and the function $f$ given by $f(|\psi(x)|,x)=|\psi(x)|^2$, Eq.~(\ref{NL-sch-eq}) admits analytic solutions \cite{WMK-2005}. For real values of $\fz$, this model has applications in the study of Bose-Einstein condensates \cite{LP-2001}. See also \cite{SCH-CW}.

The study of the NSSs for the potential~(\ref{v=d}) requires solving (\ref{nonlinear}), that for $0<x<a$ and $a<x<1$ takes the form: $\psi''+k^2\psi=\gamma f(|\psi(x)|,x)\psi(x)$. This is equivalent to
	\be
	\psi(x)=\psi_0(x)+\gamma\int_{x_0}^x \sG(x,y)f(|\psi(y)|,y)\psi(y)dy,
	\label{d-1}
	\ee
where $\psi_0(x)$ is the general solution of $\psi''+k^2\psi=0$, $\sG(x,y):=\sin[k(x-y)]/k$ is the Green's function for the latter equation, and $x_0\in(0,1)$ is arbitrary.  Using this equation to express the $\psi(y)$ appearing on its right-hand side in terms of $\psi_0$ and $\sG$ and repeating this procedure, we can obtain a perturbative expansion for $\psi(x)$ where $\gamma$ serves as the perturbation parameter. In the following we consider a homogenous nonlinearity where $f$ does not explicitly depend on $x$, i.e., $f=f(|\psi(x)|)$, and perform a first-order perturbative treatment of the NSSs of (\ref{v=d}).

First, we recall that substituting (\ref{v=d}) in (\ref{nonlinear}) is equivalent to demanding that $\psi$ is continuous at $x=a$ and that $\psi'(a^+)=\psi'(a^-)+\fz\psi(a)$, where $\psi'(a^{+/-})$ stands for the right/left derivative of $\psi$ at $x=a$. Next, we use (\ref{d-1}) to obtain a perturbative expression for the solution $\xi_k$ of (\ref{NL-sch-eq}) in the interval $[0,a)$ and use the continuity of $\psi$ at $x=0$ and the above matching condition for $\psi'$ to extend it to $[a,1]$. Finally, we demand that the result also satisfies the second equation in (\ref{condi-nonlinear-2}). This gives a pair of equations that we can solve to express $\fz$ and $\xi_k(1)$ in terms of $k,a,$ and $N_-:=\xi_k(0)$. The result is
	\begin{align}
	&\fz=2ik\left(1+\frac{\gamma f_- A}{4k^2}\right)+\cO(\gamma^2),
	\label{fz=}
	\end{align}
and $\xi_k(1)=e^{-2ika}N_-\left(1+\frac{\gamma f_- B}{4k^2}\right)+\cO(\gamma^2)$,
where $f_-:=f(|N_-|)$, $A:=e^{2ik(1-a)}+e^{2ika}-2$, and $B:=e^{2ik(1-a)}-e^{2ika}+2ik(2a-1)$. Eq.~(\ref{fz=}) is the condition under which $\psi^{(\gamma)}_{k-}$ yields a NSS. For $\gamma=0$, it reduces to $\fz=2i k$ which determines the corresponding LSS  \cite{jpa-2006,pra-2012}. Similarly we find that $\psi_{k+}^{(\gamma)}$ gives rise to a NSS provided that we enforce (\ref{fz=}) after replacing $f_-$ with $f_+:=f(|N_+|)$.

Let $r$ and $s$ denote the real and imaginary parts of $\fz$, so that $\fz=r+i s$. Noting that $f_-$ is real, we can solve (\ref{fz=}) for $\gamma f_-$ and $s$ in terms of $a$, $k$ and $r$. This gives
    \begin{align}
    &\gamma f_-\approx -\frac{k r}{\sin k \cos[k(1-2a)]},
    \label{ze1}\\
    &s\approx 2k-\left\{\frac{\cos k \cos[k(1-2a)]-1}{\sin k \cos[k(1-2a)]}\right\}\,r,
    \label{ze2}
    \end{align}
where we use $\approx$ to mean that we ignore $\cO(\gamma^2)$. Because cosine is an even function, these equations are invariant under the $\cP$-transformation: $a\to 1-a$. Therefore, we can confine our attention to the case that $a\leq\frac{1}{2}$.

\begin{figure}
    \begin{center}
    \includegraphics[scale=.65]{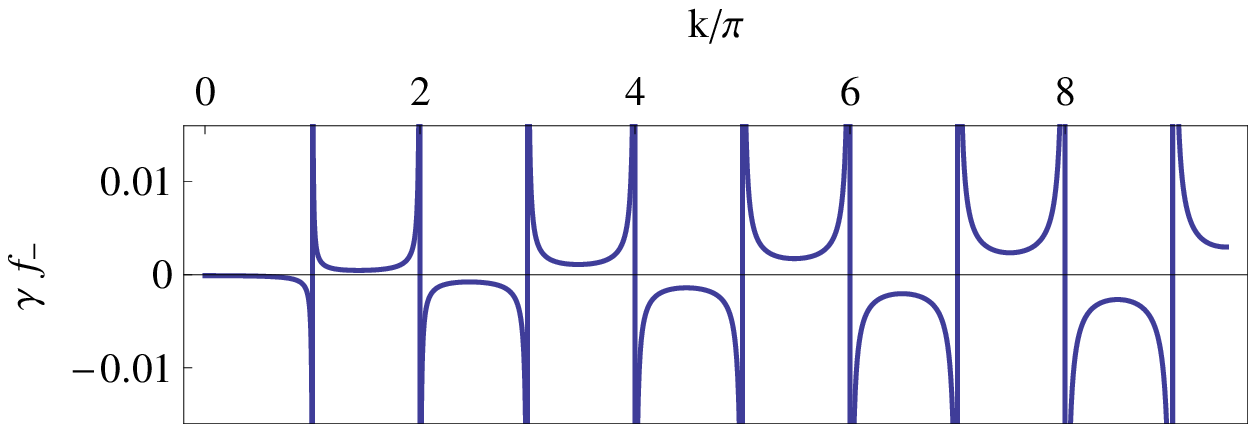}\\[6pt]
	\includegraphics[scale=.65]{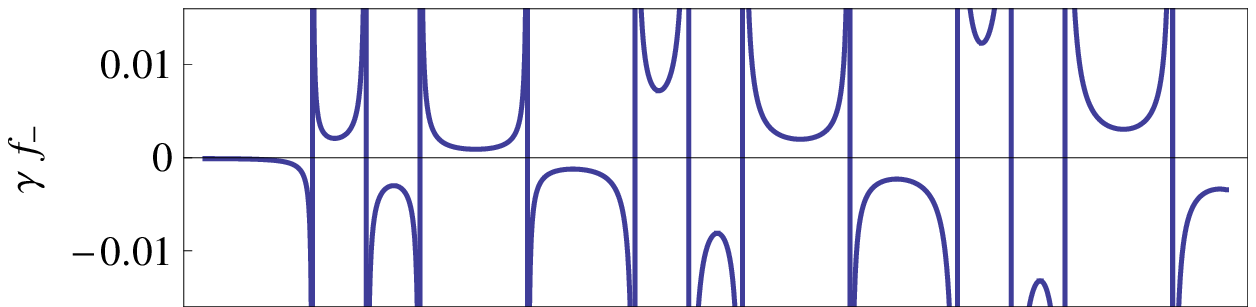}\\[6pt]
	\includegraphics[scale=.65]{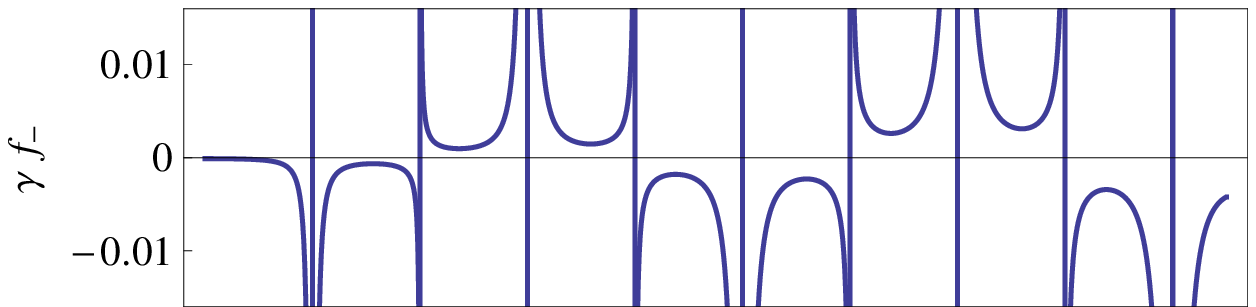}\\[6pt]
	\includegraphics[scale=.65]{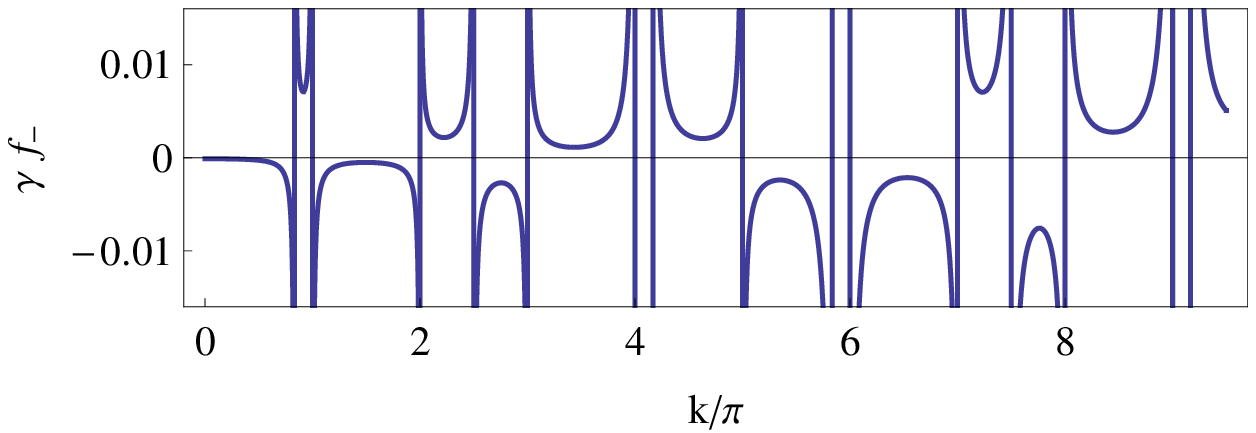}\vspace{-3mm}
    \caption{(Color online) The location of the NSSs in the $\frac{k}{\pi}\:$--$\:\gamma f_-$ plane for $r=10^{-4}$ and $a=\frac{1}{2},\frac{1}{3},\frac{1}{4},\frac{1}{5}$ respectively from top to bottom. The vertical lines are asymptotes corresponding to the $k/\pi$ values violating the condition (\ref{condi-2}).}
    \label{fig1}
    \end{center}
    \end{figure}

Eqs.~(\ref{ze1}) and (\ref{ze2}) provide a reliable description of the NSSs of (\ref{v=d}) provided that the right-hand side of (\ref{ze1}) is much smaller than $k^2$. This implies that $0<|r|\ll |\sin k \cos[k(1-2a)]|k$. In particular,
    \be
    0<|r|\ll k,
    \label{condi-1}
    \ee
and for all integers $m$,
	\be
	k\neq\left\{\begin{aligned}
	&\pi m &&\mbox{for all $a$},\\
	&\frac{\pi(m+1/2)}{1-2a} &&\mbox{for $a\neq\frac{1}{2}$}.
	\end{aligned}\right.
	\label{condi-2}
	\ee
Furthermore, (\ref{ze2}) -- (\ref{condi-2}) give $s\approx 2k$. Figure~\ref{fig1} shows the plots of $\gamma f_-$ as a function of $k$ for $r=10^{-4}$ and $a=\frac{1}{2},\frac{1}{3},\frac{1}{4},\frac{1}{5}$.

The following are some remarkable features of NSSs of the $\delta$-function potential (\ref{v=d}) that distinguishes them from their linear counterpart. Notice that they hold irrespective of the form of the nonlinearity profile $f(|\psi|)$.\\[3pt]
1.~The condition for the creation of a NSS is highly sensitive to the value of $a$.\\[3pt]
2.~Depending on the sign of $\gamma f_-$, there are specific spectral gaps for NSSs. For example, as shown in Figure~\ref{fig1}, for $\gamma f_->0$ and $a=1/3$, no NSS arises for the $k/\pi$ values in the intervals $[1.5,2]$, $[3,4]$, $[4.5,5]$, $[6,7]$, $[7.5,8]$, $[9,10]$, etc. In addition, the interval $[0,0.5]$ is forbidden for all values of $a$ whenever $\gamma f_->0$. For $\gamma f_-<0$, NSSs reside on the spectral gaps of the case $\gamma f_->0$. In particular, LSSs are continuously related to the NSSs of the case $\gamma f_-<0$.\\[3pt]
3.~There is always a minimum value of $|\gamma f_-|$ below which no NSS arises for $k>\frac{\pi}{2(1-2a)}>\frac{\pi}{2}$ if $a\neq\frac{1}{2}$ and $k>\pi$ for $a=\frac{1}{2}$. We will refer to this value of $|\gamma f_-|$ as the ``nonlinearity threshold (NT)''.

Suppose that the above model provides a description of an optical system consisting of a very thin planar slab of high-gain material. Because $s$ is proportional to the gain coefficient \cite{pra-2011a}, we can adjust the value of $s$ by controlling the pumping intensity. If $|\gamma f_-|<{\rm NT}$, the system does not lase and the incident wave does not undergo a substantial amplification regardless of how large $s$ is. Now, suppose that $|\gamma f_-|\geq {\rm NT}$. Then as we increase $s$ starting from zero, we find no amplification of the incident wave unless $s$ reaches $2k_1$, where $k_1$ is the smallest value of $k$ such that $(k,\gamma f_-)$ corresponds to a NSS. Because we can use (\ref{ze1}) to relate the values of $\gamma f_-$ and $k$, we can determine one in terms of the other. For a Kerr nonlinearity, where $\gamma f_-=\gamma|N_-|^2$, we can, in principle, employ this scheme to determine the frequency of the (incident) wave in terms of the amplitude of the transmitted wave.

The rich structure depicted in Figure~\ref{fig1} suggests other potential applications of NSSs. For example, the parameter $a$, that signifies the center of the $\delta$-function potential, can also be used as a control parameter in an experimental study of the above-mentioned frequency measurement scheme. Another possibility is to use independent frequency and intensity measurements together with the information about the location of NSSs to determine the coefficient of the Kerr and higher order nonlinearities of the medium.\\[3pt]

\noindent\emph{{Concluding Remarks}}--- In this letter we introduced the concept of a NSS for arbitrary confined nonlinearities and explored their properties for potentials having a compact support. In particular we explored in some detail NSSs of a complex $\delta$-function potential and showed that they had a much richer structure than their linear counterparts. Our results for this very simple model suggest, among other possibilities, a method for determining the frequency of an incident wave by performing an amplitude measurement.

The results we report here may be viewed as a first step toward the study of the applications of nonlinear spectral singularities in various areas of physics. This might for example lead to the discovery of the analogs of threshold lasing and antilasing for nonlinear fields such as those encountered in acoustics, Bose-Einstein condensates, fluid mechanics, and even gravity.\\[3pt]

\emph{Acknowledgments}.---  We wish to thank Aref Mostafazadeh and Ali Serpeng\"{u}zel for useful discussions. This work has been supported by  the Scientific and Technological Research Council of Turkey (T\"UB\.{I}TAK) in the framework of the project no: 110T611, and by the Turkish Academy of Sciences (T\"UBA).

\ed